\documentclass[12pt]{article}
\usepackage{color,amsmath,amssymb,exscale,psfrag,epsfig}
\usepackage{cite,color,url}
\usepackage{pdfpages}
\usepackage{graphicx}
\usepackage[colorlinks=true
,urlcolor=blue
,anchorcolor=blue
,citecolor=blue
,filecolor=blue
,linkcolor=blue
,menucolor=blue
,linktocpage=true
,pdfproducer=medialab
]{hyperref}
\input epsf
\newcommand{\m}[1]{\marginpar{{\tiny *}} }

\newcommand{\Dslash}{{\not \!\!D}}

\def\bea{\begin{eqnarray}}
\def\eea{\end{eqnarray}}

\newcommand\f[2]{\frac{#1}{#2}}
\catcode`\@=11
\def\lsim{\mathrel{\mathpalette\@versim<}}
\def\gsim{\mathrel{\mathpalette\@versim>}}
\def\@versim#1#2{\vcenter{\offinterlineskip
\ialign{$\m@th#1\hfil##\hfil$\crcr#2\crcr\sim\crcr } }}
\catcode`\@=12
\catcode`@=12
\begin{document}
\thispagestyle{empty}
\begin{flushright}
ICAS 014/16
\end{flushright}
\vspace{0.3in}
\begin{center}
{\Large \bf A 750 GeV graviton and the Higgs as a pNGB} \\
\vspace{0.5in}
{\bf Ezequiel Alvarez$^{(a,b)}$,
Leandro Da Rold$^{(c)}$,\\[1ex]
Javier Mazzitelli$^{(a)}$
and
Alejandro Szynkman$^{(d)}$}
\vspace{0.2in} \\
{\sl $^{(a)}$ International Center for Advanced Studies (ICAS), UNSAM, Campus Miguelete \\
25 de Mayo y Francia, (1650) Buenos Aires, Argentina
}
\\[1ex]
{\sl $^{(b)}$ International Center for Theoretical Physics (ICTP), Strada Costiera, 11, Trieste, Italy 
}
\\[1ex]
{\sl $^{(c)}$ Centro At\'omico Bariloche, Instituto Balseiro and CONICET \\
Av. Bustillo 9500, 8400, S. C. de Bariloche, Argentina
}
\\[1ex]
{\sl $^{(d)}$ IFLP, Dpto. de F\'{\i}sica, CONICET, UNLP \\ 
C.C. 67, 1900 La Plata, Argentina} \\
\end{center}
\vspace{0.5in}

\begin{abstract}
We study the diphoton excess at 750 GeV reported by ATLAS and CMS, by assuming that it corresponds to a new spin-two resonant state. We model this state as a massive graviton in a two-site model. We show that the very stringent bounds from $VV$ final states can be evaded naturally by considering that the Higgs is a pseudo Nambu-Goldstone boson. In this case the couplings of the graviton to the longitudinal electroweak gauge bosons can be parametrically suppressed. On the other hand, partial compositeness allows to suppress the leptonic channels.
We compute loop-induced contributions to the graviton couplings by the presence of the SM third generation of quarks and composite partners of the SM fermions and obtain that they are not important.
We find that the diphoton signal and the experimental constraints from other decay channels can be reproduced in a large and natural region of the parameter space of the theory. 
\end{abstract}

\vspace*{20mm}
\noindent {\footnotesize E-mail:
{\tt \href{mailto:sequi@df.uba.ar}{sequi@df.uba.ar},
\href{mailto:jmazzitelli@unsam.edu.ar}{jmazzitelli@unsam.edu.ar},
\href{mailto:daroldl@cab.cnea.gov.ar}{daroldl@cab.cnea.gov.ar},\\
\href{mailto:szynkman@fisica.unlp.edu.ar}{szynkman@fisica.unlp.edu.ar}}}


\newpage
\section{Introduction}
During the last couple of decades high energy physicists have been busy discovering the last particles predicted by the Standard Model (SM) and looking for New Physics (NP) beyond the SM.   In this search for NP, experimentalists have been mainly guided by theorists who would point out where NP was likely to appear, according to many kind of theoretical and general arguments.  In addition, experimentalists have also looked for NP in other not so popular channels as it is for instance the diphoton channel.  Even though we still need more data to determine if this is a discovery or a fluctuation, a diphoton excess at 750 GeV has been detected by ATLAS \cite{Aaboud:2016tru} and CMS \cite{Khachatryan:2016hje} at a combined level of $\sim 3-4$ standard deviations and, nowadays, is an open window from which NP could be envisaged.

Contrary to thinking that there is very little information on this hypothetical resonance, we should think that the modest diphoton signal  is accompanied by a huge lack of signal in all other channels, therefore providing abundant and difficult data to be reproduced by any model that attempts to explain the full data-set.  

In fact, since the very first theoretical papers \cite{Franceschini:2015kwy,Ellis:2015oso,Gupta:2015zzs} aiming to understand what could be the nature of this resonance, many other papers have addressed the challenging task of fitting the whole data in, for instance and not being exhaustive, the framework of supersymmetry \cite{Benakli:2016ybe,Cohen:2016kuf,Ellwanger:2016qax}, Two Higgs Doublet Models \cite{Bizot:2015qqo,Gopalakrishna:2016tku,Han:2016bvl}, bound states \cite{Kats:2016kuz,Han:2016pab,Aparicio:2016iwr}, scalars~\cite{Csaki:2016kqr,Cox:2015ckc} and gravitons from extra dimensions~\cite{Arun:2015ubr,Giddings:2016sfr,Falkowski:2016glr,Han:2015cty,Kim:2015ksf,Buckley:2016mbr,Martini:2016ahj,Geng:2016xin,Sanz:2016auj,Hewett:2016omf,Carmona:2016jhr}, among many others. One of the main obstacles in many of the proposals is related to the lack of signal in other channels, specifically in the searches of $t\bar t$, $ZZ$ and $WW$ final states, which in generic models have large branching ratios.

In this work, we address the possibility that the new resonance could be a spin-two massive state. As recently shown in Refs.~\cite{Falkowski:2016glr,Hewett:2016omf}, obtaining a spin two resonance parametrically lighter than the other states of the new sector, as well as avoiding the bounds from $VV$ final states, are the most challenging issues. There are several possiblities to lower its mass \cite{Falkowski:2016glr,Hewett:2016omf}. In this work we show that, if the Higgs is a pseudo Nambu-Goldstone boson (pNGB), the coupling between the spin-two state and the longitudinal electroweak (EW) vectors can be suppressed, without tuning. Although the suppression is mild, it is enough to avoid the bounds in the problematic channels. We will compute the production cross-section of the diphoton signal, as well as the other possible final states. We will show that the signal can be reproduced, passing successfully all the experimental bounds, in a large and natural region of the parameter space. Since a pNGB Higgs arising from a strongly interacting sector offers a compelling solution addressing the hierarchy problem and the origin of electroweak symmetry breaking (EWSB), we find it very interesting that the same mechanism can also lead to a natural explanation of the diphoton excess.

A natural framework for describing a massive spin-two state as well as a pNGB Higgs is to consider that there is a new strongly interacting sector beyond the SM. The diphoton resonance corresponds to the lowest lying spin-two massive state of this sector, and it can be associated with a resonance of the graviton. Assuming that the sector has a global symmetry spontaneously broken to a smaller subgroup by the strong dynamics at the TeV scale, the Higgs emerges as the Nambu-Goldstone boson (NGB) of this breaking. It has been shown that the interactions with the SM states can trigger EWSB dynamically~\cite{Agashe:2004rs}, and that EW precision tests can be passed at the price of a moderate tuning~\cite{Agashe:2005dk,Carena:2007ua,Grojean:2013qca}. The duality of strongly interacting theories in four dimensions and weakly coupled theories in extra dimensions offers a perturbative description~\cite{Maldacena:1997re,Randall:1999ee}. An alternative and attractive possibility is to consider a 2-site model~\cite{Contino:2006nn}, that can be related with the discretization of an extra dimensional theory, keeping only the lowest lying level of resonances. We will work within this framework, that besides its simplicity, does not require a rigid relation between the masses of the spin-one and spin-two states.

In addition to the massive graviton, there will also be resonances associated to the SM fermions and gauge bosons. Although these states do not play a direct role in the phenomenology of the diphoton resonance, they can lead to corrections of the couplings at loop level.  We will compute the one-loop contributions induced by the SM third generation of quarks and the heavy fermions. We will show that, for the region of the parameter space that favours the diphoton signal, these corrections are small, and the tree-level results give an accurate description of the phenomenology. We will also show that the model generically predicts that the final states $VV$ should be accessible in the near future.

The paper is organized as follows. In section~\ref{sec:model} we describe the two-site model and show the relevant couplings. Section~\ref{sec:pheno} contains the predictions for the signal and the region of the parameter space that can reproduce it. In section~\ref{sec:loop} we compute the one-loop corrections to the couplings by the presence of the SM third generation of quarks and the new fermions. In section~\ref{sec:predictions} we discuss some predictions of the model, and finally we conclude in section~\ref{sec:conclusions}. We leave some details of the one-loop calculation for the Appendix.

\section{Model}\label{sec:model}
\begin{figure}[t] 
\begin{center}
\includegraphics[width=0.6\textwidth]{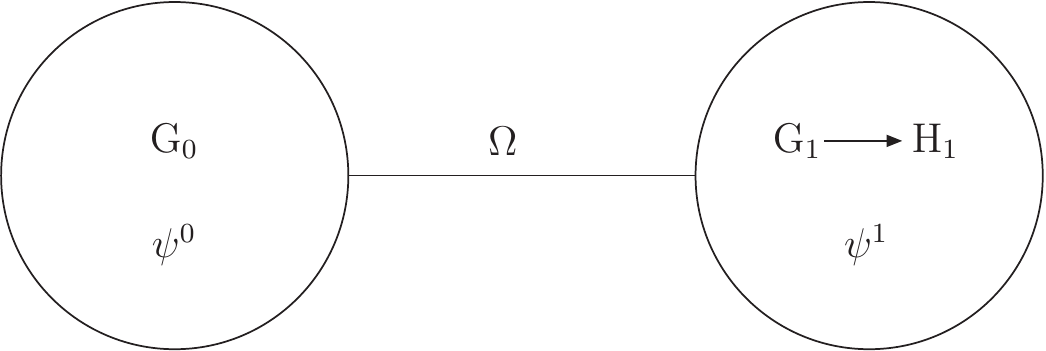}
\end{center}
\caption{\small Moose diagram of the two site theory describing the model. G$_0$ and G$_1$ are the gauge symmetries on sites 0 and 1. The interactions on site-1 spontaneously break G$_1$ down to H$_1$. There is a set of link fields $\Omega$ for gauge and gravitational interactions.}
\label{fig-moose}
\end{figure}

We consider a model with two sites: site-0 containing {\it elementary} fields and site-1 containing the first level of {\it composite} resonances of a strongly interacting sector, see Fig.~\ref{fig-moose}. Site-0 has the same gauge symmetry and fermion fields as the SM, plus a massless graviton and no Higgs. Site-1 is similar to the SM, it has a larger gauge symmetry in the EW sector to include the custodial symmetry and to deliver a pNGB Higgs, it also contains a graviton resonance and several multiplets of fermions. The fermions are in representations of the gauge symmetry group on site-1, and there is a full multiplet of {\it composite} fermions for each multiplet of fermions of the {\it elementary} sector.\footnote{It is also possible to consider more than one {\it composite} fermion for each {\it elementary} one~\cite{Contino:2006qr,Csaki:2008zd,Andres:2015oqa}.} We will describe the Higgs and the extended EW symmetry in section~\ref{sec-higgs-sector}. All the couplings on site-1 are assumed to be larger than the SM couplings: $g_1\gg g_{SM}$, although still perturbative: $g_1\ll4\pi$. 
Both sites are connected by link fields, we have denoted them collectively as $\Omega$ in the moose diagram. There is a link field $Y_{10}$ that transforms under general coordinate transformations on site-0 and 1. Calling these general coordinate transformations $F_0$ and $F_1$: $Y_{10}\to F^{-1}_1\circ Y_{10}\circ F_0$. $Y_{10}$, being a mapping from site 1 to site 0, allows to compare fields living on the different sites. We will work in the unitary gauge for the link field $Y_{10}$, where the map corresponds to the identity~\cite{ArkaniHamed:2002sp}. There are also link fields $U$ connecting the gauge symmetry groups of both sites, we will describe them in detail in section~\ref{sec-higgs-sector}.

An important ingredient that we want to include in the model is partial compositeness of the SM fermions, that can be realized if there are linear interactions between the elementary fields and the operators of the strongly interacting sector~\cite{Contino:2004vy}. In the two site model partial compositeness can be achieved by introducing linear mixing between the fermions on site-0 and site-1, with the corresponding factors of the link fields connecting them, to maintain the symmetries of the theory. We will consider the case where all the SM fermions acquire mass by partial compositeness.

In the unitary gauge the Lagrangian is as in Ref.~\cite{Contino:2006nn}, except for the fermion embedding and the Higgs sector that will be specified later:
\begin{align}
&{\cal L}={\cal L}_0+{\cal L}_1+{\cal L}_{\rm mix} \ , \\
&{\cal L}_j={\cal L}_j^{\rm matter}+\sqrt{-G_j}2M_j^2R(G_j) + \dots\ , \label{eqLi}\\
&{\cal L}_{\rm mix}={\cal L}_{\rm mix}^{\rm matter}+{\cal L}_{\rm mix}^{\rm grav} \ ,
\end{align}
with $G^j_{\mu\nu}$ the metric in site $j$. $M_j$ is the scale of the gravitational interactions in each side, for site-0 it is of order $M_{\rm Pl}$, whereas for site-1 it is of order TeV. The dots in Eq.~(\ref{eqLi}) are present to allow for more terms and fields, as for example a cosmological constant and a dilaton field.

The interactions with the graviton arise from the minimal coupling with the metric, to linear order in the graviton fields
\begin{equation}
{\cal L}\supset \sum_{j=0,1}X^j_{\mu\nu}T_j^{\mu\nu} \ ,
\end{equation}
where we have split the metrics in the Minkowski term and a gravitational fluctuation: $G^j_{\mu\nu}=\eta_{\mu\nu}+X^j_{\mu\nu}$, and the energy-momentum tensor on each site is defined as usual:
\begin{align}
T_j^{\mu\nu}=&-\frac{1}{g_j^2}F_j^{\mu\rho}F_{j\rho}^{\nu}+\eta^{\mu\nu}\frac{1}{4g_j^2}F_j^{\rho\sigma}F_{\rho\sigma}^j+\frac{i}{2}\bar\psi_j(\gamma^\mu D_j^\nu+\gamma^\nu D_j^\mu)\psi_j-\eta^{\mu\nu}\bar\psi_j(i\Dslash_j-m_j)\psi_j +\dots 
\end{align}
The dots stand for the contribution from scalar fields as the Higgs, that will be shown in section~\ref{sec-higgs-sector}. 

\subsection{Gauge and Higgs sectors}\label{sec-higgs-sector}
As has been stated by several authors, the nature of the Higgs has deep implications for the phenomenology of the massive graviton, because for composite Higgs, the graviton decays copiously to longitudinal $W$'s and $Z$'s, generically surpassing the bounds from LHC. This is related to the fact that the Higgs is usually considered as a composite state, completely localized on site-1. We will consider in this section the case of a pNGB Higgs, that besides from leading naturally to a light Higgs, can also give a suppression in the decay of the massive graviton to longitudinal gauge bosons. We will show that, without tuning, this suppression is enough to open a wide range of the parameter space of the model.

For a concrete example we will consider a pNGB Higgs arising from SO(5)/SO(4), however the analysis below is in general independent of the specific pattern of symmetry breaking, as long as the unbroken subgroup H$_1$ contains the SM gauge group. The gauge symmetry group on site-1 will be G$_1$=SO(5)$\times$U(1)$_X$, with SO(5)$\supset$SU(2)$_L\times$SU(2)$_R$ and hypercharge identified as $Y=T^{3R}+X$, as usual.\footnote{There is also an SU(3)$_c$ that we will not describe in detail because it does not play a crucial role in the analysis below.} We assume that the strong dynamics break G$_1$ to H$_1$=SO(4)$\times$U(1)$_X$, delivering a multiplet of NGB transforming as ${\bf4}_0$ of H$_1$.
For later convenience, we will consider a spurious extension of the gauge symmetry on site-0: G$_0$=SO(5)$\times$U(1)$_X$. For that purpose we will introduce non-dynamical fields on site-0, such that the {\it elementary} fields fill complete representations of the extended symmetry group. 

There are two non-linear sigma fields in the EW sector of the theory. There is a scalar field $U_A=e^{i\sqrt{2}\Pi_A/f_A}$ that transforms as $U_A \to g_0 U_A g_1^\dagger$, with $g_{0,1} \in$ G$_{0,1}$. $U_A$ parameterizes the breaking G$_0\times$G$_1$/G$_{0+1}$, with $\Pi_A=\Pi_A^rT^r$ and $T^r$ the broken generators. There is another scalar field $U_1=e^{i\sqrt{2}\Pi_1/f_1}$, that transforms as $U_1\to g_1 U_1h_1^\dagger$, with $g_1\in$ G$_1$ and $h_1\in$ H$_1$. $U_1$ parameterizes the spontaneous breaking G$_1\to$H$_1$ by the interactions on site-1, with $\Pi_1=\Pi_1^{\hat a}T^{\hat a}$. We will label with $a$ the unbroken generators of H$_1$ and with $\hat a$ the generators of G$_1$/H$_1$. The scales $f_A$ and $f_1$ breaking the symmetries are taken of order TeV. The Lagrangian contains the following kinetic terms:
\begin{equation}\label{kinetic-ngb}
{\cal L}\supset \frac{f_A^2}{4}\sqrt{-G_0}G^0_{\mu\nu}(D^\mu U_A)^\dagger D^\nu U_A +\frac{f_1^2}{4}\sqrt{-G_1}G^1_{\mu\nu}\sum_{\hat a} d^{\hat a\mu}d^{\hat a\nu} \ .
\end{equation}
The covariant derivative $D_\mu U_A$ and the symbol $d_{\mu}$ are defined by:
\begin{align}
& D_\mu U_A=\partial_\mu U_A-iA_\mu^0 U_A+iU_A A_\mu^1 \ ,
\\
& i U_A^\dagger D_\mu U_A=d_{\mu}+e_{\mu} \ , \qquad d_{\mu}=d_{\mu}^{\hat a} T^{\hat a} \ , \qquad e_{\mu}=e_{2\mu}^a T^a \ .
\end{align}

The kinetic terms of Eq.~(\ref{kinetic-ngb}) mix the NGB fields $\Pi_A$ and $\Pi_1$ with the gauge fields $A_\mu^0$ and $A_\mu^1$. The mixing terms are:
\begin{equation}\label{kinetic-ngb2}
{\cal L}\supset \frac{f_A}{\sqrt{2}}\sum_{r=a,\hat a}(A_\mu^{0r}-A_\mu^{1r})\partial^\mu\Pi^r_A +
 \frac{f_1}{\sqrt{2}}\sum_{\hat a}A_\mu^{1\hat a}\partial^\mu\Pi^{\hat a}_1 \ .
\end{equation}
Taking into account that the gauge fields of G$_0$/H$_0$ are not dynamical, the unitary gauge in this sector, defined by the absence of mixing, corresponds to $\Pi_A^a=0$, $\Pi_A^{\hat a}=f_h/f_A\Pi^{\hat a}$ and $\Pi_1^{\hat a}=f_h/f_1\Pi^{\hat a}$, with:
\begin{equation}
\frac{1}{f_h^2}=\frac{1}{f_A^2}+\frac{1}{f_1^2} \ . 
\end{equation}
$\Pi^{\hat a}$ is the only physical scalar, with a decay constant $f_h$. It can be associated with the Higgs field, and due to the explicit breaking of the symmetries by the mixing terms it becomes a pNGB with a potential induced at one-loop level. As has been shown in many papers, this potential can trigger EWSB dynamically, leading to a realistic model for $\xi\equiv\sin^2(v/f_h)\sim0.1$~\cite{Agashe:2004rs}. There are several possibilities for the representations of the fermions, that can account for different effects in the protection of the couplings~\cite{Agashe:2006at} and in the tuning of the potential~\cite{Panico:2012uw,Carena:2014ria}. Since the phenomenology that we will study is rather independent of this choice, we will not specify these representations. We will only consider a specific case in section~\ref{sec:loop}, where we study the one-loop correction to the couplings by the presence of heavy fermions. As we will show, in general the correction is small, thus we postpone the description of a possible representation to section~\ref{sec:loop}.

\subsection{Interactions in the mass basis}\label{sec-mass-basis}
To study the phenomenology of the graviton resonance we describe below the rotations that lead to the mass basis, as well as the graviton couplings in this basis. We will not consider in this section the mixing effects generated by EWSB, that will require new rotations of order $g v/m_1$, with $m_1\sim$TeV the masses of the states on site-1.

${\cal L}_{\rm mix}^{\rm grav}$ breaks the symmetry  of general coordinate transformations on site-1 and 2 to the diagonal subgroup, with a linear combination of the graviton fields $X^0_{\mu\nu}$ and $X^1_{\mu\nu}$ that becomes massive. The term leading to the graviton mass in unitary gauge is~\cite{ArkaniHamed:2002sp}:
\begin{align}
&{\cal L}_{\rm mix}^{\rm grav}=-\frac{f_X^4}{2}\sqrt{-G_0}(K_{\mu\rho}K_{\nu\sigma}-K_{\mu\nu}K_{\rho\sigma})(K^{\mu\rho}K^{\nu\sigma}-K^{\mu\nu}K^{\rho\sigma}) \ , \\
&K_{\mu\rho}=G^0_{\mu\rho}-G^1_{\mu\rho} \ ,
\end{align}
with $f_X\sim$TeV. 

Similarly, ${\cal L}_{\rm mix}^{\rm matter}$ breaks G$_0\times$G$_1\to$G$_{0+1}$, generating a mass for $(A_\mu^0-A_\mu^1)$, and the spontaneous breaking G$_1\to$H$_1$ gives an extra contribution to the fields in the coset G$_1/$H$_1$:
\begin{equation}\label{kinetic-ngb3}
{\cal L}\supset \frac{f_A^2}{4}\sum_{r=a,\hat a}(A_\mu^{0r}-A_\mu^{1r})^2+
 \frac{f_1^2}{4}\sum_{\hat a}(A_\mu^{1\hat a})^2 \ .
\end{equation}

Let us describe now the fermion sector. For simplicity we assume that there is one {\it composite} fermion in a given representation of G$_1$ for each {\it elementary} fermion in a given representation of the EW group. In section~\ref{sec:loop} we show an example for the quarks. The fermions on site-1 are vector-like, with masses $m_{\psi_1}=g_\psi f_1\sim$TeV, generated by the strong dynamics, where $g_\psi$ is a composite coupling $\sim{\cal O}(1)$ and we have chosen $f_1$ as the normalization scale for $m_{\psi_1}$. In the simplest description all the couplings on site-1 are of the same size $g_\psi\simeq g_1$, however to avoid fine-tuning of the Higgs potential a smaller $g_\psi$ is preferred for the third generation, whereas to avoid large corrections to the EW observables a larger $g_1$ is preferred. We will not elaborate more on this issue, but the reader should keep in mind that, for the quarks of the third generation, $g_\psi\lesssim g_1$. ${\cal L}_{\rm mix}^{\rm matter}$ mixes the chiral fermions on site-0 with the corresponding partner on site-1:
\begin{equation}\label{eq-mpsi}
{\cal L}\supset \sum_{\psi}\Delta_\psi \bar\psi^0\psi^1-\sum_{\psi^1}m_{\psi_1}\bar\psi^1\psi^1 \ ,
\end{equation}
where the sum in the first term is over all the fermions on site-0 and their corresponding partners, and in the second term over all the fermions on site-1. Since G$_1$ is broken to H$_1$, the representations of the fermions on site-1: ${\bf r}_{\text{G}_1}$, decompose under H$_1$ as ${\bf r}_{\text{G}_1}\simeq\oplus_\alpha {\bf r}^\alpha_{\text{H}_1}$, where ${\bf r}^\alpha_{\text{H}_1}$ are the irreducible representations of H$_1$ contained in ${\bf r}_{\text{G}_1}$. Thus in general there can be a different $m_{\psi_1}$ for each $\alpha$.

The kinetic term of the graviton and gauge fields can be canonically normalized by field redefinitions: $A^j_\mu\to g_j A^j_\mu$ and $G^j_{\mu\nu}\to G^j_{\mu\nu}/M_j$. By taking the elementary couplings to zero (as well as the fermionic mixing) we obtain massless elementary fields and massive composite fields: $m_{A_1^a}=g_1f_A/\sqrt{2}$, $m_{A_1^{\hat a}}=g_1(f_A^2+f_1^2)^{1/2}/\sqrt{2}$, $m_{X_1}=f_X^2/M_1$ and $m_{\psi_1}$.\footnote{Up to corrections, as in the case of spontaneous symmetry breaking in the strongly coupled sector, for example in models with the Higgs being a pNGB.} The mixing between the fields in sites 0 and 1 can be diagonalized by a simple rotation (before EWSB)
\begin{align}
&\Phi=c_\Phi \Phi_0+s_\Phi \Phi_1 \ , \qquad \Phi^*=-s_\Phi \Phi_0+c_\Phi \Phi_1 \ , \qquad t_\Phi=\frac{s_\Phi}{c_\Phi} \ , \label{eq-rotation} \\
& t_A=\frac{g_0}{g_1} \ , \qquad t_\psi=\frac{\Delta}{m_{\psi_1}} \ , \qquad t_X=\frac{M_1}{M_0} \ ,
\end{align}
where $\Phi_j$ is any of the fields in the original Lagrangian, $\Phi$ denotes the massless field and $\Phi^*$ the massive one, with mass: $m_{\Phi^*}=m_{\Phi_1}\sqrt{1+t_\Phi^2}$. The names of the variables $s_\Phi,\ c_\Phi$ and $t_\Phi$ stand for the trigonometric functions $\sin\theta_\Phi,\ \cos\theta_\Phi$ and $\tan\theta_\Phi$. Since there are different gauge symmetry groups, the ratio $t_A$ and the scale $f_1$ can be different for different groups, however for simplicity in this paper we will consider these quantities to be the same for all the groups.
As discussed at the beginning of section~\ref{sec-higgs-sector}, only a subset of the fields on site-0 are  dynamical. Thus in general there are fields on site-1 that do not mix with any field on site-0. These fields are usually called custodians and have masses $m_{\Phi_1}$, as explained above Eq.~(\ref{eq-rotation}).

After EWSB the would-be massless fermions acquire masses of order: $m_{\psi}^{SM}\simeq s_{\psi_L}s_{\psi_R}v g_\psi$, with $g_\psi=m_{\psi_1}/f_1$ setting the mass scale of the vector-like fermions on site-1, Eq.~(\ref{eq-mpsi}). Since the masses of the SM fermions are proportional to the Left- and Right-handed mixing, the top quark requires $s_{q}, s_{t}\sim0.5-1$. The small mass of the the light quarks can be obtained by choosing at least one of the chiral mixing small. In the rest of the paper we will assume that the mixing of all the fermions is small, except for the quarks of the third generation. For this reason we will not consider the effect of the light fermions in the phenomenology of the massive graviton.
In the simplest model the {\it elementary} fermions mix with one multiplet {\it composite} fermion only, thus the mixing in the quark sector is parametrized by $s_q$, $s_t$ and $s_b$. In this case $s_b$ gives the suppression for the small bottom mass. In some models, as MCHM$_5$, $q_L$ has to mix with two {\it composite} fermions at least, to generate the quark masses. In this case there is an extra mixing: $s_{q'}$, that controlls the bottom mass. We will assume that $s_{q'}$ is small and we will neglect its effect in the following.

The mixing leaves unbroken the diagonal subgroup of the original symmetries, thus there is a set of massless gauge fields $A_\mu$, as well as a massless graviton $X_{\mu\nu}$. These gauge and graviton fields interact with universal couplings: $g^{-2}=g_0^{-2}+g_1^{-2}$ and $M^{-2}=M_0^{-2}+M_1^{-2}$, respectively.

By rotating the fields to the mass basis we can write schematically the interactions with the massive graviton as
\begin{equation}\label{eq:Lgrav}
{\cal L}\supset \sum_\Phi \tilde C_\Phi X^*_{\mu\nu}T^{\mu\nu}(\Phi) \ .
\end{equation}
The energy-momentum tensor of the Higgs, $T^{\mu\nu}(H)$, can be taken as in the linear case~\cite{Falkowski:2016glr}. The non-linearity can give small corrections and take place in three particle decays.
We obtain the following couplings for the interactions of Eq.~(\ref{eq:Lgrav}):
\begin{align}
& \tilde C_{A}=-\frac{s_A^2c_X}{M_1}+\frac{c_A^2s_X}{M_0} \ ,
\qquad \tilde C_{\psi}=\frac{s_\psi^2c_X}{M_1}-\frac{c_\psi^2s_X}{M_0} \ , \\
& \tilde C_{A^*}=-\frac{c_A^2c_X}{M_1}+\frac{s_A^2s_X}{M_0}  \ ,
\qquad \tilde C_{\psi^*}=\frac{c_\psi^2c_X}{M_1}-\frac{s_\psi^2s_X}{M_0}  \ , \\
& \tilde C_{A-A^*}=2s_A c_A\left(\frac{c_X}{M_1}-\frac{s_X}{M_0}\right)  \ ,
\qquad \tilde C_{\psi-\psi^*}=-2s_\psi c_\psi\left(\frac{c_X}{M_1}-\frac{s_X}{M_0}\right) \ , \label{eq:XtoKKandSM}\\
& \tilde C_H=\frac{c_X}{M_1}\frac{f_h^2}{f_1^2}-\frac{s_X}{M_0}\frac{f_h^2}{f_A^2} \ , \label{eq-X-H}
\end{align}
where $\tilde C_{\Phi-\Phi^*}$ are interactions involving a massless and a massive field, obtained after the diagonalization. For very small mixing ($s_\Phi\ll 1$) we obtain the well known universal behaviour of the couplings between a heavy resonance and a current of massless fields, with a coupling that to leading order is independent of the mixing of the light state: $\tilde C\simeq\pm s_X/M_0$. 

For $M_1\ll M_0$ and mixing angle not too small, the couplings with the massive graviton are dominated by the first term: $\tilde C_\Phi\simeq \pm c_X s_\Phi^2/M_1$. For later convenience we will define a dimensionless coupling $C_\Phi$ by factorizing the scale of gravity on site-1:
\begin{equation}
\tilde C_\Phi\equiv C_\Phi/M_1 \ .
\end{equation}

For fields on site-1 that do not mix with with site-0, the coupling with the massive graviton is given by $c_X/M_1$, thus it is completely fixed by the gravity scale on site-1. This is the case for example for the custodians, as well as for the Higgs in models where it is fully localized on site-1. As can be seen from Eq.~(\ref{eq-X-H}), the case of a pNGB Higgs is different, below we describe it briefly. We define a mixing angle for the pNGB Higgs as
\begin{equation}
t_H=\frac{f_A}{f_1}, \qquad s_H=\frac{f_h}{f_1}, \qquad c_H=\frac{f_h}{f_A} \ ,
\end{equation}
where again $s_H,\ c_H$ and $t_H$ stand for $\sin\theta_H,\ \cos\theta_H$ and $\tan\theta_H$. For $M_1\ll M_0$, Eq.~(\ref{eq-X-H}) is dominated by the first term: $C_H\simeq s_H^2$, thus the coupling is modulated by $s_H^2=f_h^2/f_1^2$. For example, for $f_A=f_1$ one obtains $C_H=1/2$, leading to a suppression in the Higgs coupling, that can play an interesting role evading the experimental constraints in $X^*\to VV$, as we will show in section~\ref{sec:pheno}.

Naturality in our model prefers values of $s_H^2\simeq 0.2-0.8$, however in the next section we will allow $s_H$ to depart from this value, reaching also values near the extremes: $0<C_H<1$. By doing so one can effectively describe other models that could lead to different Higgs couplings.

\section{Phenomenology}\label{sec:pheno}
Using the above described model we can easily understand the phenomenology of this scenario by parameterizing the graviton production and its branching ratios through the free variables in the model, namely $s_A,\ s_{q},\ s_{t},\ s_{b},\ C_H$ and $M_1$.  In the following we study tree-level phenomenology and leave one-loop effects for the next section.

At tree level, using MadGraph \cite{Alwall:2014hca} with PDF NN23LO1  we have
\begin{eqnarray}
\sigma(pp\to X^*) &=& \left\{ 
\begin{array}{cc} 
\left(\frac{2\text{ TeV}}{M_1}\right)^2 167.4 s_A^4  \mbox{pb} & 13 \mbox{TeV} \\
\left(\frac{2\text{ TeV}}{M_1}\right)^2 37.88 s_A^4  \mbox{pb} & 8 \mbox{TeV,} 
\end{array} 
\right.
\end{eqnarray}
where we have assumed a QCD $k$-factor $k=1.6$ \cite{Falkowski:2016glr}.  We have verified that including tree level $b \bar b \to X^*$ production accounts at most to a 5-10\% of the total cross-section production, therefore for simplicity we do not include this contribution in this discussion.

The formulae for the width of the graviton to the different particles can be found elsewhere \cite{Falkowski:2016glr}, however we quote here the relevant ones for the discussion that follows,
\begin{eqnarray}
\Gamma(X^* \to f\bar f) \!\!\!&=&\!\!\! \frac{N_c m_X^3}{320 \pi M_1^2} (1-4r_f)^{3/2} \left( (|C_{f_L}|^2 + |C_{f_R}|^2) \left( 1-\frac{2 r_f}{3} \right) + Re(C_{f_L} C_{f_R}^*)  \frac{20 r_f}{3} \right)\!, \hspace*{0.7cm}
 \label{ff} \\ 
\Gamma(X^* \to ZZ ) \!\!\!&=&\!\!\!  \frac{ m_X^3}{80 \pi M_1^2} (1-4r_Z)^{1/2} \bigg( |C_{Z}|^2 + \frac{|C_H|^2}{12} + \frac{r_Z}{3} \Big( 3|C_H|^2 - 20 Re(C_H C_{Z}^*) - 9 |C_{Z}|^2 \Big) \nonumber \\
&& + \frac{2 r_Z^2}{3} \Big(  7|C_H|^2 +10  Re(C_H C_{Z}^*) + 9 |C_{Z}|^2  \Big) \bigg), \label{zz}\\ 
\Gamma(X^* \to Z\gamma) \!\!\!&=&\!\!\! \frac{m_X^3}{160\pi M_1^2} \tan(2\theta_W)^2 |C_Z-C_\gamma|^2  (1-r_Z)^3 \left( 1 + \frac{r_Z}{2} + \frac{r_Z^2}{6} \right), \label{za} \\
\Gamma(X^* \to \gamma\gamma) \!\!\!&=&\!\!\! \frac{|C_\gamma|^2 m_X^3}{80 \pi M_1^2}, \label{aa} \\ 
\Gamma(X^* \to HH) \!\!\!&=&\!\!\! \frac{|C_H|^2 m_X^3}{960 \pi M_1^2}  (1-4r_H)^{5/2} , \label{hh}  
\end{eqnarray}
whereas $\Gamma(X^* \to WW )=2\Gamma(X^* \to ZZ )$ replacing $m_Z\to m_W$ and $C_Z \to C_W$.
Here $r_i = m_i^2/m_X^2$.
Notice that, since we have chosen a universal mixing for all the gauge groups, $C_\gamma=C_Z\simeq - s_A^2$ and the $X^* \to Z\gamma$ is not allowed at tree-level. However this is only a simplified picture, and the decay $X^* \to Z\gamma$ can be open if the mixing differs for the different groups of the EW sector.

We look for the region in parameter space that can simultaneously satisfy the bounds in Table \ref{limits}. Notice that we are analyzing all the 8 TeV limits using the original constraints to the cross-section, and not the relationship between their widths and  the diphoton width, as usual.  In fact, the latter is only equivalent to the former in the region where the diphoton cross-section is satisfied in its central value and, therefore, produces a bias when analyzing the phenomenology outside this region, which is important to understand the overall behaviour of the model.

\begin{table}
\[
\begin{array}{|l|c|c|}
\hline
\mbox{observable} &  \mbox{allowed} & \mbox{collider energy} \\
\hline
\sigma(pp\to X^*) \times \mbox{BR}(\gamma\gamma) & 5.5 \pm 1.5 \mbox{ fb}  & \mbox{13 TeV} \\
\sigma(pp\to X^*) \times \mbox{BR}(\gamma\gamma) & < 1.5 \mbox{ fb}  & \mbox{8 TeV} \\
\sigma(pp\to X^*) \times \mbox{BR}(jj) & < 2.5 \mbox{ pb}  & \mbox{8 TeV} \\
\sigma(pp\to X^*) \times \mbox{BR}(h h)& < 39\ \mbox{ fb} & 8 \mbox{ TeV} \\
\sigma(pp\to X^*) \times \mbox{BR}(ZZ)& < 12\ \mbox{ fb} &  8 \mbox{ TeV} \\
\sigma(pp\to X^*) \times \mbox{BR}(WW)& < 40\ \mbox{ fb}  & 8 \mbox{ TeV}  \\
\sigma(pp\to X^*) \times \mbox{BR}(t\bar t)& < 450\ \mbox{fb} & 8 \mbox{ TeV} \\
\sigma(pp\to X^*) \times \mbox{BR}(b \bar b)& < 1\ \mbox{pb} & 8 \mbox{ TeV} \\
\hline
\end{array} 
\]
\caption{\small Bounds on different channels to be satisfied by the model \cite{Franceschini:2015kwy}.}
\label{limits}
\end{table}

Observe that all graviton couplings to the particles in Table \ref{limits} have an upper bound of $1/M_1$ in this model.  We will consider that all the scales on site-1 are of the same order. Since the fermion resonances have a lower bound $m_{\psi_1}\gtrsim1$ TeV from direct searches~\cite{CMS:vwa}, and the vector resonances have a lower bound $m_{A_1}\gtrsim2-3$ TeV from EWPT~\cite{Barbieri:2004qk} and direct searches, we will take $M_1 \gtrsim 1$ TeV. For this lower bound on $M_1$, the fermion couplings modulated by $s_{q},\ s_{t}$ and $s_{b}$ will not have a dominant role in determining the allowed parameter space, since their limit in Table \ref{limits} is in general not saturated, even for the lowest allowed value for $M_1$ and maximal mixing. Observe, however, that the branching ratios to fermions may be dominant, even though with little change when scanning in other variables of the model. This is numerically verified below.

We are therefore left with the relevant parameters $s_A,\ C_H$ and $M_1$.  We analyze their impact on the observables in the following paragraphs.  A good approach is to first understand the effects of $s_A$ and $C_H$ independently of $M_1$, and then analyze the effect of $M_1$ using what is learned from $s_A$ and $C_H$.  Notice that $M_1$ does not affect the branching ratios, but only the production cross-section. 

\begin{figure}
\begin{center}
\includegraphics[width=0.47\textwidth]{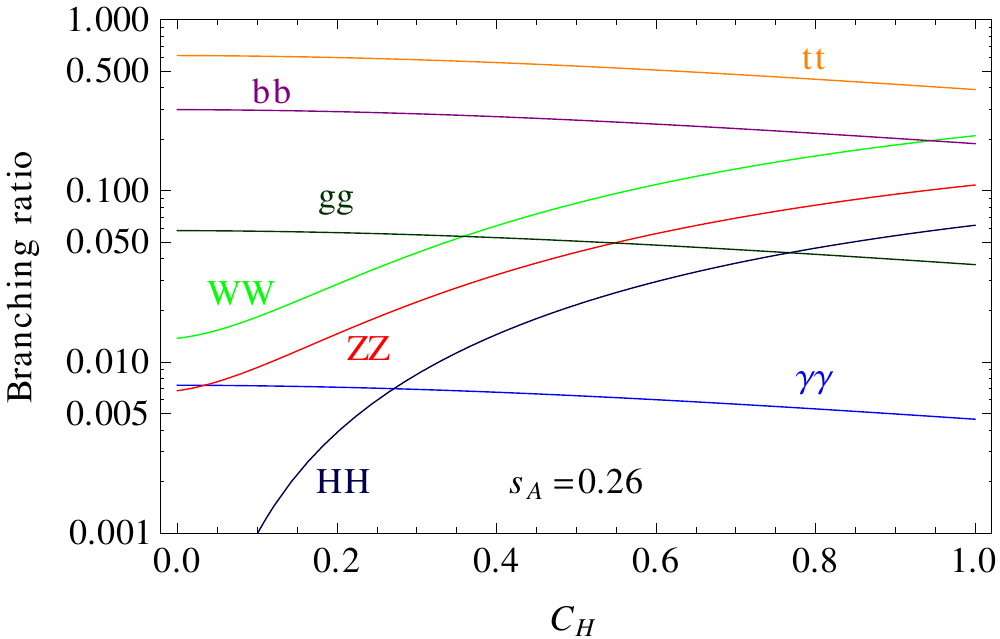}~
\includegraphics[width=0.47\textwidth]{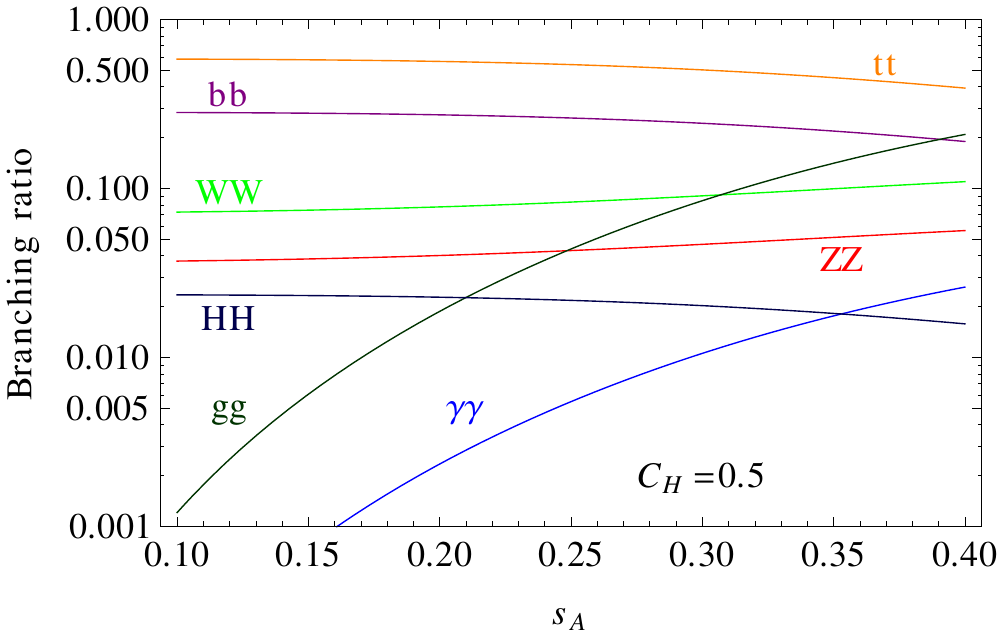}
\caption{
\small
Graviton branching ratios as a function of the two left relevant variables in the model.  See text to understand the behaviour from Eqs.~(\ref{zz}) and (\ref{aa}).  
}
\label{brs}
\end{center}
\end{figure}

As it can be seen from the above formulae, an increase in $s_A$ increases the cross-section production and also the width to $\gamma\gamma,\ ZZ$ and $WW$.  However, the widths to $ZZ$ and $WW$ also have a dependence on $C_H$ coming from its longitudinal polarization.   On the other hand, the width to $HH$ is only driven by $C_H$, however the small numerical factor in front of Eq.~(\ref{hh}) makes this final state not of main relevance for constraining the parameter space.   Since the width to fermions is independent of $s_A$ and $C_H$, then the widths to $\gamma\gamma$, $ZZ$ and $WW$ are the ones that provide the most relevant modifications to the graviton branching ratios, which is one of the main objects to determine the allowed parameter space.  At last, observe that the $jj$ bound in Table \ref{limits} does not impose any limit in parameter space since we consider universal couplings $C_g=C_\gamma$, whereas on the other hand we may expect some light tension with the $\gamma\gamma$ limit for 8 TeV.  We do not consider the $\gamma\gamma$ limit at 8 TeV in the following, and at the end of this section we verify that only a 10\% of the selected parameter space is discarded by this requirement.  

\begin{figure}[t!]
\begin{center}
\includegraphics[width=0.47\textwidth]{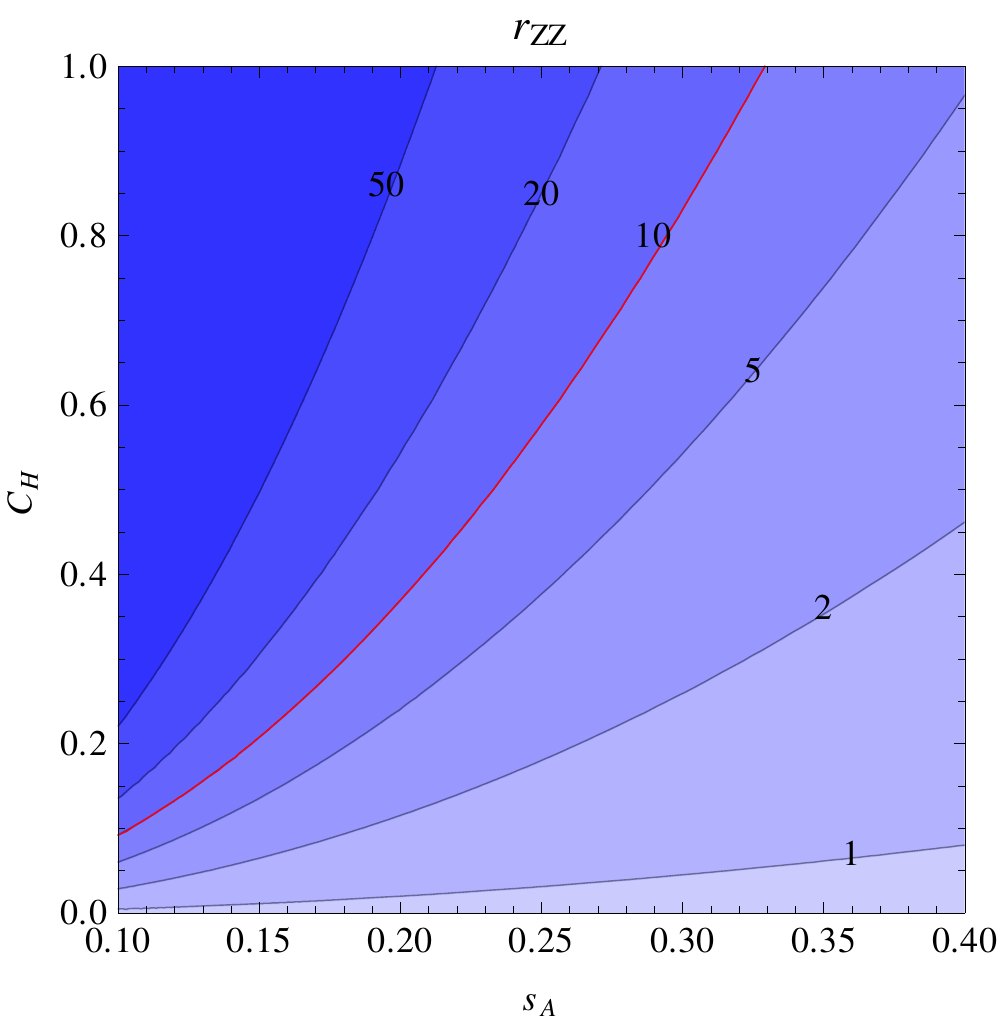}~
\includegraphics[width=0.45\textwidth]{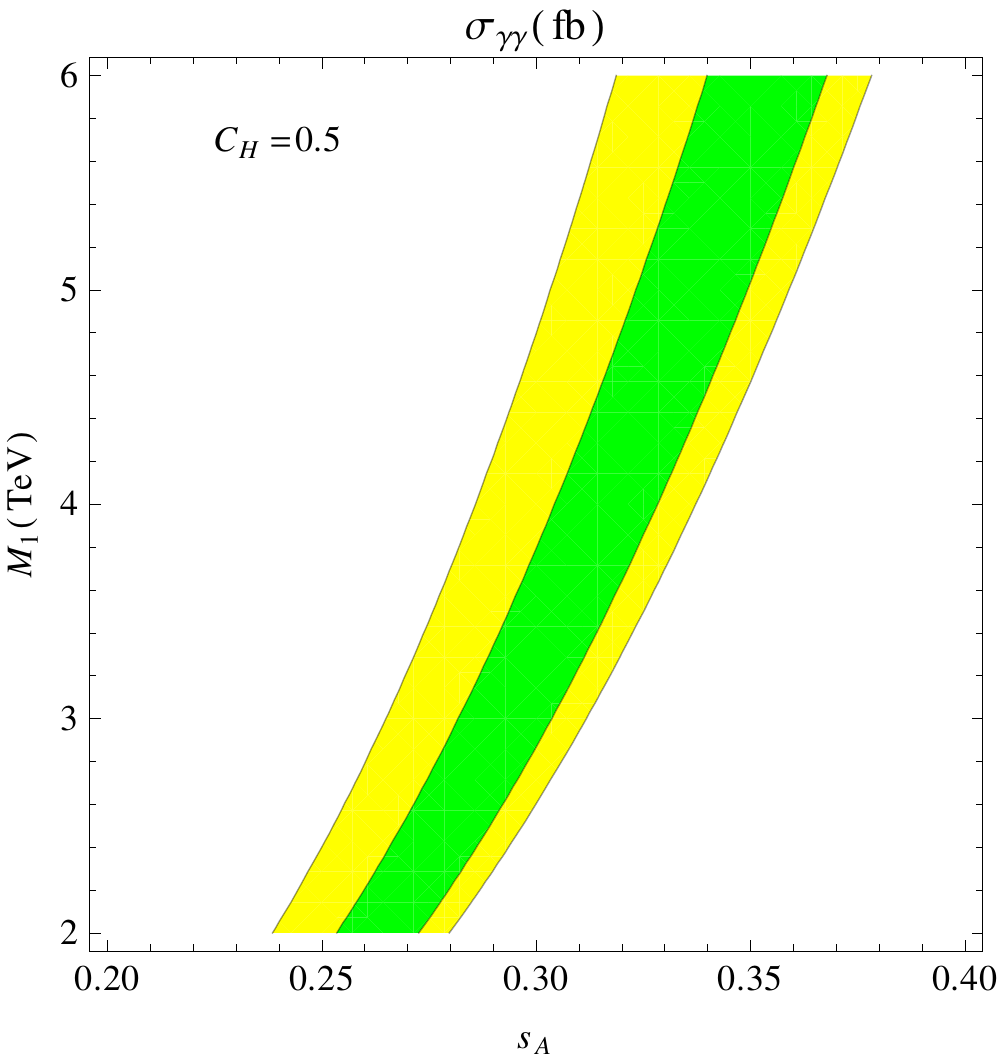}
\caption{
\small
Left: ratio between the $ZZ$ and $\gamma\gamma$ partial decay widths, $r_{ZZ}$, in the $s_A - C_H$ plane.
The $r_{ZZ}=10$ curve corresponding to the current experimental constraints --assuming the $\sigma\times $BR$(\gamma\gamma)$ is fixed to its central value-- is indicated in red.
Right: total cross section for $pp \to X^* \to \gamma\gamma$ at $13\text{ TeV}$, in the $s_A - M_1$ plane. The green and yellow bands indicate the $1\sigma$ and $2\sigma$ regions for the experimentally measured excess.  As explained in text, larger values of $M_1$ (smaller production cross-sections) need larger values of $s_A$ (larger BR$(\gamma\gamma)$).
}
\label{contour-ratio}
\end{center}
\end{figure}

Since data indicates a non zero $\mbox{BR}(\gamma\gamma)$ but has strong upper bounds on $\mbox{BR}(ZZ,WW)$, then we need to bound $\Gamma(X^* \to ZZ,WW)$ when compared to $\Gamma(X^* \to \gamma\gamma)$.  In fact, it is easily seen from Eqs.~(\ref{zz}) and (\ref{aa}) that if either $s_A$ increases or $C_H$ decreases then the ratio of widths $r_{ZZ} = \Gamma_{ZZ}/\Gamma_{\gamma\gamma}$ is reduced.  To illustrate this discussion we plot in Fig.~\ref{brs} the branching ratios for the graviton as a function of $s_A$ with $C_H$ fixed, and vice-verse, and in Fig.~\ref{contour-ratio}a the contour lines of $r_{ZZ}$.
The red curve in Fig.~\ref{contour-ratio}a represents the experimental limit on this ratio (provided that the $\gamma\gamma$ central value is reproduced); the region to the right of this curve is allowed by the $ZZ$ constraints.
These plots as well as other results in this section use a benchmark point defined as $s_A=0.26$, $s_{q}=0.7$, $s_{t}=0.8$, $s_{b}=0.3$, $C_H=0.5$ and $M_1=2$ TeV: when any of these variables is missing, then it is taken from this benchmark point.
In particular, this point reproduces the diphoton signal and fulfills all other experimental constraints.
We foresee that playing with $s_A$ and $C_H$ will allow us to find regions in parameter space which satisfy better or worse the constraints in Table \ref{limits}.  In particular, one should note that by modifying $C_H$ one addresses directly the limits on $\sigma \times$BR$(ZZ,WW)$, whereas BR$(\gamma\gamma)$ is only modified indirectly.

Once that the joint impact of $s_A$ and $C_H$ on the observables has been understood, we can proceed to study the impact of $M_1$ on the phenomenology.

Since increasing $M_1$ decreases the absolute value of all couplings, then the production cross-section decreases without affecting the branching ratios.  Therefore, one needs a shift in another variable to obtain a larger branching ratio to photons in order to keep $\sigma \times $BR$(\gamma\gamma)$ within the allowed values.  According to the previous discussion, this could be achieved by increasing $s_A$ and/or decreasing $C_H$; we can see from Fig.~\ref{brs} that the former is more sensitive.  To illustrate this dependence we plot in Fig.~\ref{contour-ratio}b the $\sigma\times$BR$(\gamma\gamma)$ allowed parameter space in the $s_A$-$M_1$ plane for $C_H$ fixed and the benchmark point defined above.  The dependence in $C_H$ of this figure is mild.

Finally, using the previous analysis, we can understand how the allowed parameter space behaves as a function of the relevant variables $s_A$, $C_H$ and $M_1$.  One could proceed, for instance, as follows.  For a given value of $C_H$ in the model, Fig.~\ref{contour-ratio}a --which is independent of $M_1$-- tells us which are the allowed values for $s_A$ when one assumes the diphoton cross-section fixed in its central value.  For any of these allowed values, one can find out in Fig.~\ref{contour-ratio}b --slightly adapted to the corresponding $C_H$-- which are the values of $M_1$ that will give positive solutions for the model and data.  

\begin{figure}
\begin{center}
\includegraphics[width=0.47\textwidth]{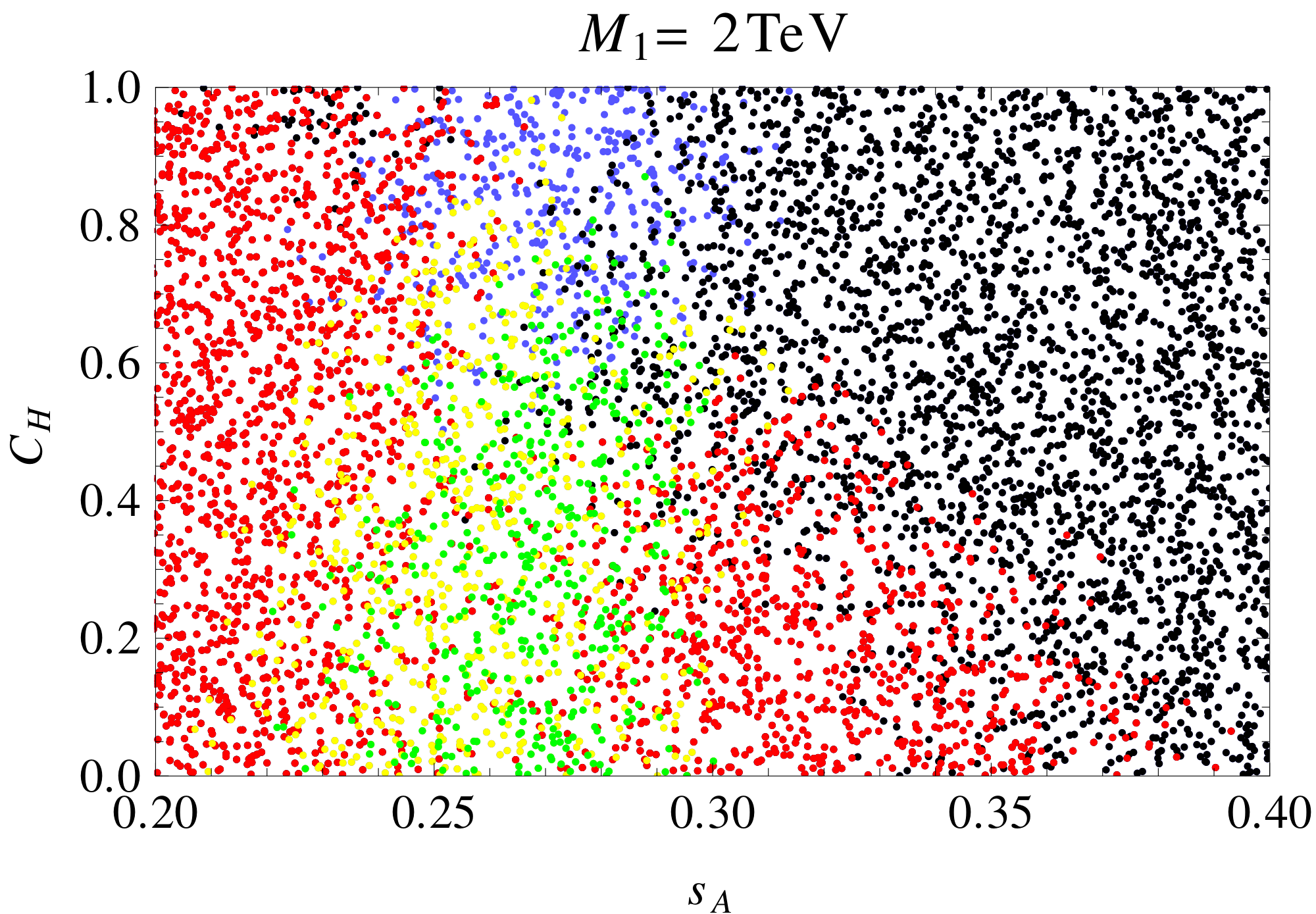}~
\includegraphics[width=0.47\textwidth]{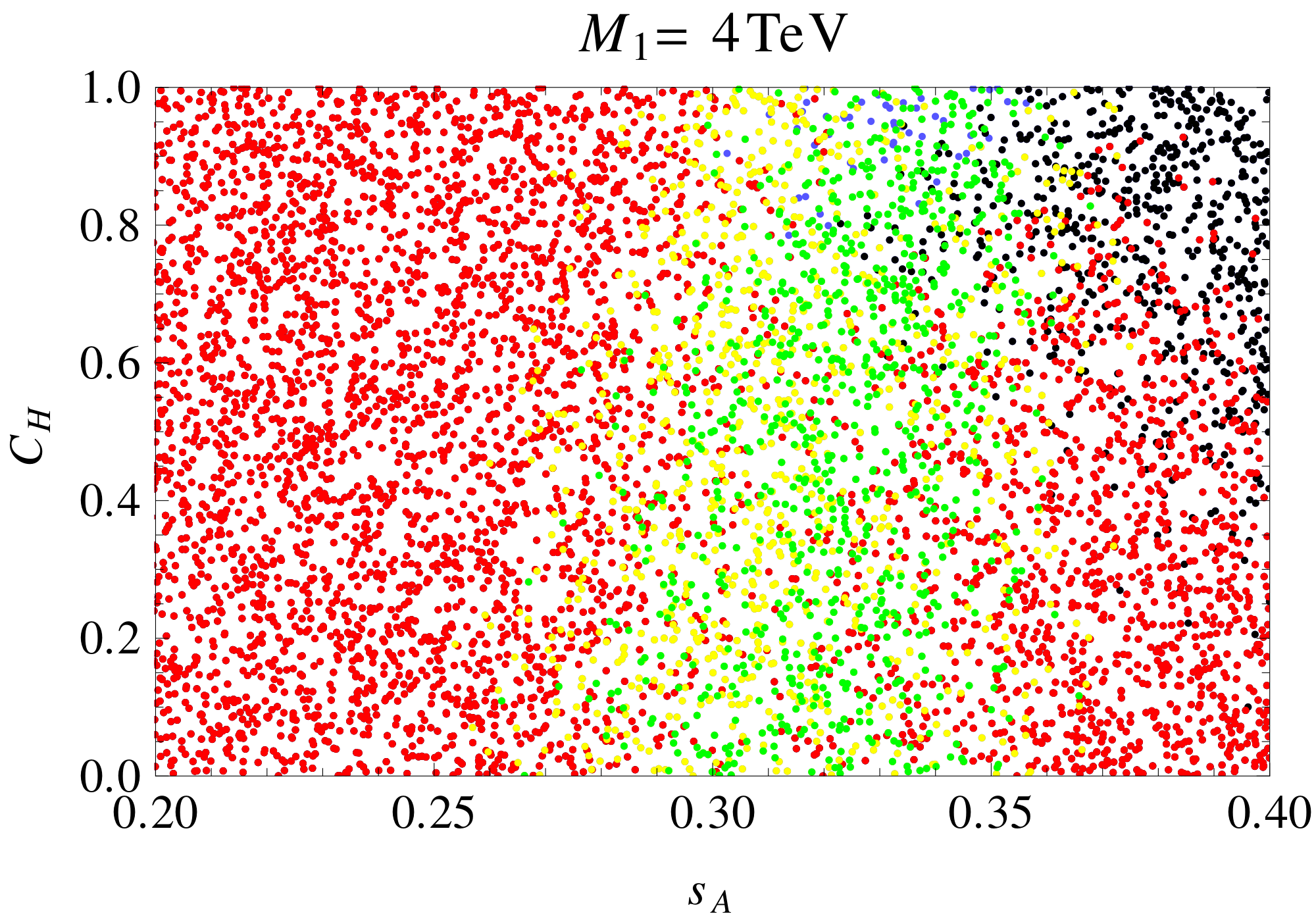}
\caption{
\small
The values of $C_H$ and $s_A$ for the points of the parameter space simulated (specified in the main text), for $M_1=2$ and $4\text{ TeV}$.
Green points: reproduce the $\gamma\gamma$ signal within $1\sigma$ and fulfill all the experimental constraints.
Yellow points: reproduce the $\gamma\gamma$ signal within $2\sigma$ and fulfill all the experimental constraints.
Red points: do not reproduce the $\gamma\gamma$ signal, but still fulfill all other experimental constraints.
Blue points: reproduce the $\gamma\gamma$ signal within $1\sigma$ or $2\sigma$, but do not fulfill all other experimental constraints.
Black points: do not reproduce the $\gamma\gamma$ signal, and neither fulfill all other experimental constraints.
}
\label{scatter}
\end{center}
\end{figure}

To verify this expected behaviour we have randomly scanned the parameter space in all variables ($s_A \in (0.1,0.4),\ s_{q}\in (0.5,0.95),\ s_{t}\in (0.5,0.95),\ s_{b}\in (0.1,0.9)$ and $C_H\in (0,1)$ for different values of $M_1$) and plot in a $s_A$-$C_H$ plane which points pass different types of bounds.  The results are shown in Fig.~\ref{scatter} for the two cases $M_1=2$ and $4$ TeV (see the caption for the color coding).  We see that for larger values of $C_H$ one needs larger $M_1$, being this one of the main results in this work.

There are also many other features to be understood from the plots in Fig.~\ref{scatter}.  First, since there is little overlap of colors, then the dependence on the not plotted variables ($s_{q}$, $s_{t}$ and $s_{b}$) is very small, as predicted from the analysis above.  We also see that there is an imaginary line dividing the red, yellow and green region from the black and blue region, which goes to larger values of $C_H$ as $M_1$ increases: this is because the larger is $M_1$ (the smaller the production cross-section) then the more relaxed are the constraints on BR($ZZ,WW$), which can now increase as $C_H$ increases.  The shape of this line would have been different (more likely to the contours in Fig.~\ref{contour-ratio}a) if we would have used constraints on $r_{ZZ,WW}$ instead of absolute constraints as depicted in Table \ref{limits}.  One could also understand why the green and blue region slightly bends to larger $s_A$ as $C_H$ increases: as $C_H$ increases the BR$(\gamma\gamma)$ is indirectly reduced and then $s_A$ should slightly increase to compensate and keep constant BR$(\gamma\gamma)$.

At this point we return to the constraint on $\sigma \times $BR$(\gamma\gamma)$ at 8 TeV in Table \ref{limits}.  When requiring this last constraint on the allowed points in parameter space (green points in Fig.~\ref{scatter}) we obtain that only 10\% of the points get rejected.  These are located in the bottom right part of the green points, as expected.  

We should also quote that the total width of the graviton in term of its mass for the green points in Fig.~\ref{scatter} is roughly $0.1\%$ and $0.03\%$ for $M_1=2$ and $4$ TeV, respectively.

We have thus worked out the phenomenology of the model and obtained points in parameter space that can satisfy all the requirements.  Along this section we have understood that in order to have $C_H=1$ one needs to increase the scale $M_1$,
but that a large region of the parameter space is allowed for $C_H<1$, also for lower values of $M_1$.
As discussed in section \ref{sec:model}, these values of $C_H$ can be naturally obtained if the Higgs is a pNGB.

\section{Stability upon loop corrections}\label{sec:loop}
In order to evaluate the stability of the tree-level predictions presented in the previous section, we consider here the fermion loop contributions to the production and decay of the massive graviton to photons and gluons.

In the first place, we include the contributions coming from bottom and top-quark loops.
This leads to the following effective couplings,
\begin{eqnarray}
\label{eqn:effCoup}
C_{\gamma}^{\text{eff}} &=&
C_{\gamma} + \f{\alpha}{2\pi}
\left[
s_{q}^2
\left(
\f{1}{9}A_G(\tau_b)+\f{4}{9}A_G(\tau_t)
\right)
+ \f{4}{9} s_{t}^2 A_G(\tau_t)
+ \f{1}{9} s_{b}^2 A_G(\tau_b)
\right]\,,
\\
C_{g}^{\text{eff}} &=&
C_{g} + \f{\alpha_S}{2\pi}\f{1}{6}
\left[
s_{q}^2
\left(
A_G(\tau_b)+A_G(\tau_t)
\right)
+ s_{t}^2 A_G(\tau_t)
+ s_{b}^2 A_G(\tau_b)
\right]\,,\label{eqn:effCoup2}
\end{eqnarray}
where $\tau_i = 4 m_i^2 / m_X^2$, and the loop function $A_G(\tau)$ takes the form \cite{Geng:2016xin}
\begin{eqnarray}
A_G(\tau) &=& -\f{1}{12}\bigg[
-\f{9}{4}\tau(\tau+2) [2 \tanh^{-1}(\sqrt{1-\tau}) - i\pi]^2
\\
&+& 3(5\tau+4)\sqrt{1-\tau}[2 \tanh^{-1}(\sqrt{1-\tau}) - i\pi]
- 39\tau + 12\ln \tau - 35 - 12\ln 4 \bigg]\nonumber
\end{eqnarray}
for $\tau<1$.
All of the other decay channels are evaluated at tree-level.
Notice that these loop corrections are proportional to the squared mixing angle, and thus the contributions from light fermions can be safely neglected.

\begin{figure}
\begin{center}
\includegraphics[width=0.47\textwidth]{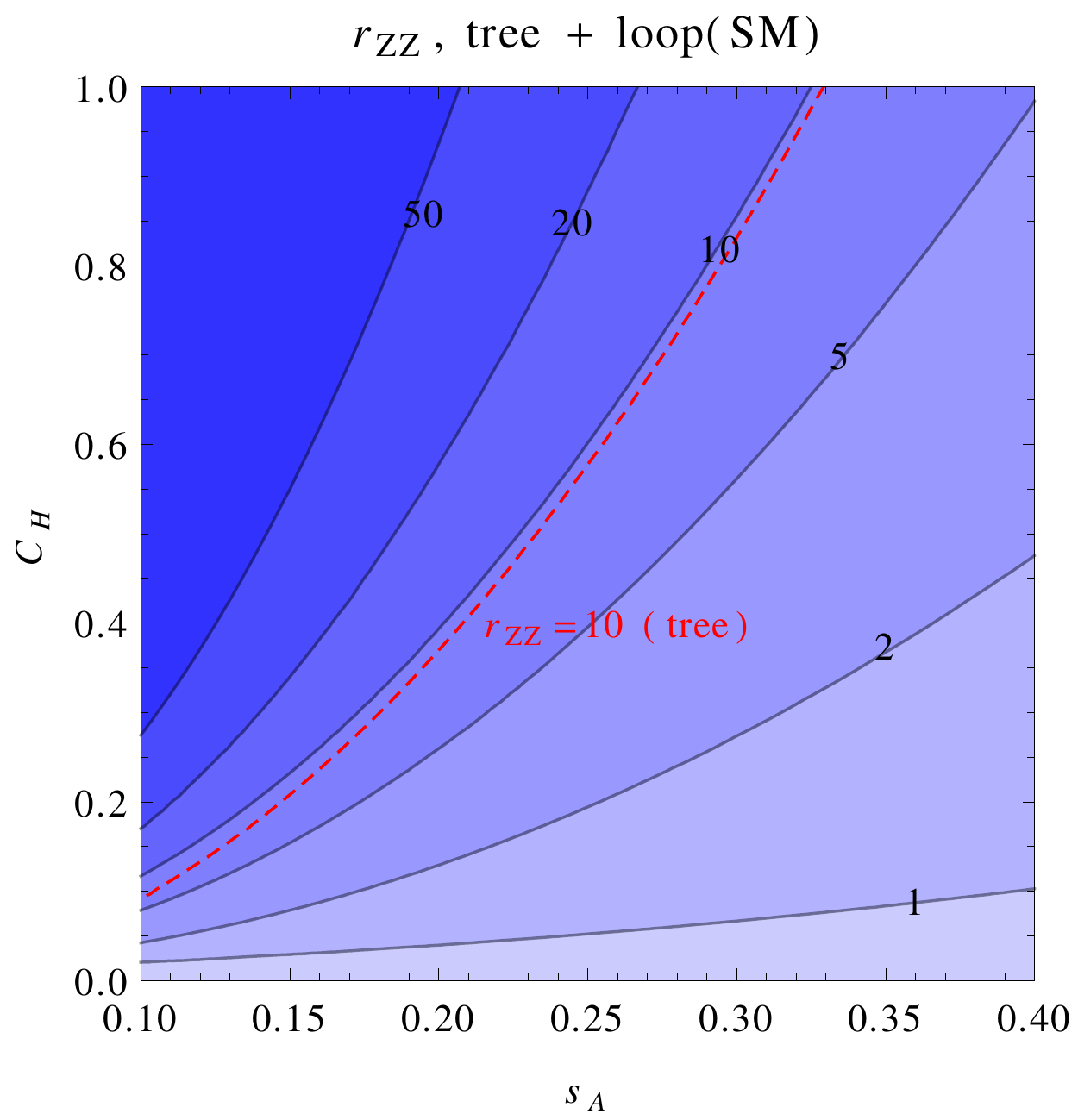}~
\includegraphics[width=0.45\textwidth]{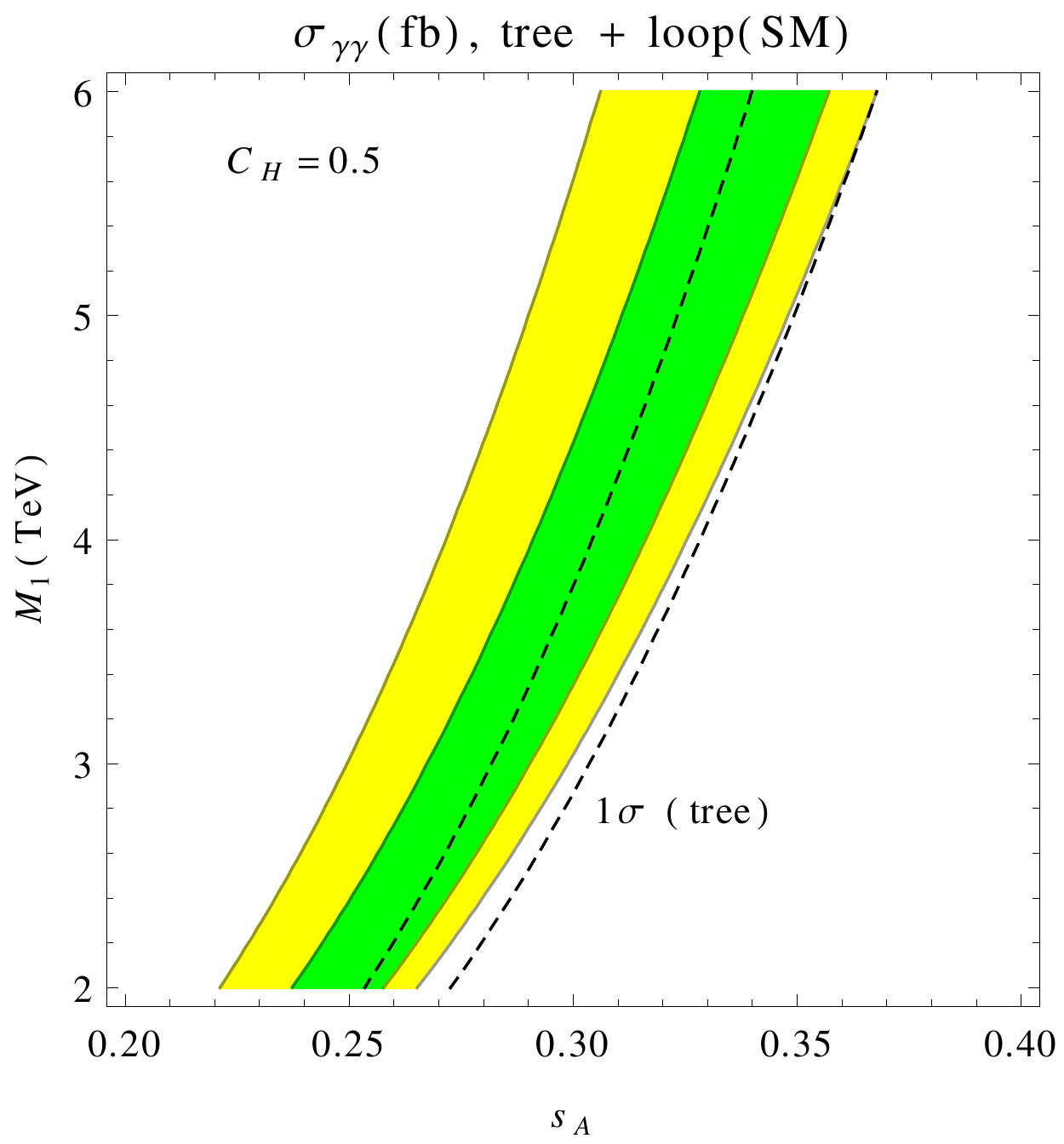}
\caption{
\small
Same quantities as the ones shown at tree level in Fig.~\ref{contour-ratio}, including the loop induced contributions from SM fermions.
Left: the red-dashed line indicates the $r_{ZZ}=10$ curve of Fig.~\ref{contour-ratio}a.
Right: the dashed curves indicate the $1\sigma$ region of Fig.~\ref{contour-ratio}b.
}
\label{contour-ratio-loop}
\end{center}
\end{figure}

The effect of the loop induced contributions coming from the SM quarks turns out to be relatively small.
This is illustrated in Fig.~\ref{contour-ratio-loop}, where the ratio $r_{ZZ}$ and the cross section $\sigma_{\gamma\gamma}$ are shown in the plane $s_A - C_H$ and $s_A - M_1$ respectively, as it was shown before at tree-level in Fig.~\ref{contour-ratio}.
In order to facilitate the comparison, the relevant tree-level curves corresponding to $r_{ZZ} = 10$ and the $1\sigma$ region for $\sigma_{\gamma\gamma}$ are also plotted.

As it can be seen from Fig.~\ref{contour-ratio-loop}, the effect of the loop contributions on the ratio $r_{ZZ}$ is rather small, finding only moderate deviations from the tree-level behaviour for all the values in the $s_A - C_H$ plane.
Of course, in this ratio the loop effects are only included for the denominator, but in view of the small size of the corrections we do not expect large deviations even if the loop contributions were included for the $ZZ$ decay width.
The effect on the $pp \to X^* \to \gamma\gamma$ cross section is also under control, as we always find an overlap between the regions that reproduce the experimentally measured excess within $1\sigma$. 

\begin{figure}
\begin{center}
\includegraphics[width=0.47\textwidth]{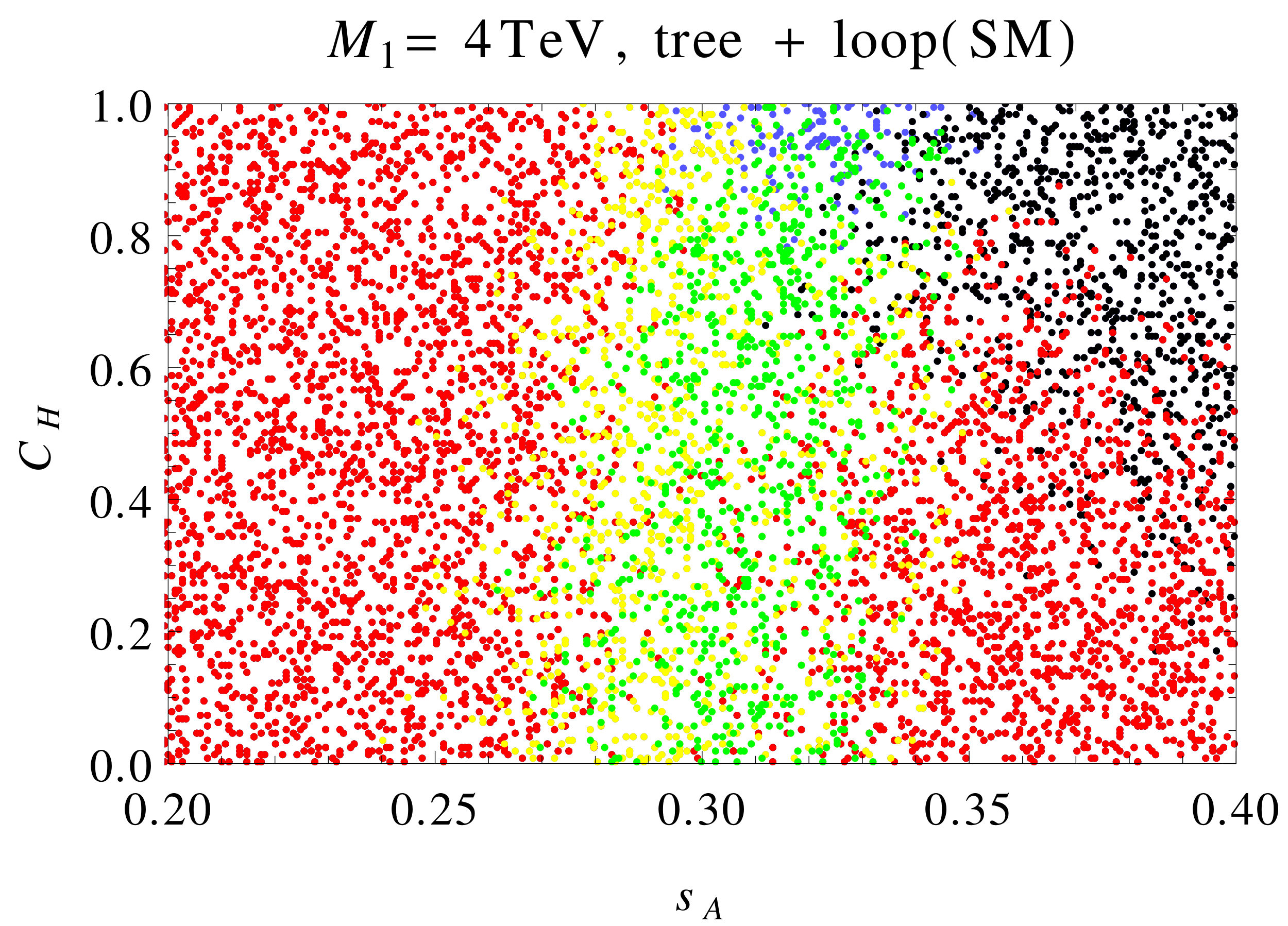}
\caption{
\small
The values of $C_H$ and $s_A$ for the points of the parameter space simulated (specified in the main text), for $M_1=4\text{ TeV}$, including the effects from bottom and top-quark loops in the $\gamma\gamma$ and $gg$ couplings.
The color coding is the same as in Fig.~\ref{scatter}.
}
\label{scatterLoop}
\end{center}
\end{figure}

Given that the two main constraints for our model, i.e. reproducing the $13\text{ TeV}$ diphoton signal and fulfilling the $8\text{ TeV}$ $ZZ$ limits, depend essentially on the variables shown in Fig.~\ref{contour-ratio-loop}, we do not expect to have large modifications on the allowed regions of the parameter space.
Nevertheless, we performed another scan on the parameters, this time including the contributions of the SM fermions in the loop described above.
As can be seen from Fig.~\ref{scatterLoop}, the results we obtained are compatible with the ones presented in section \ref{sec:pheno}.
The main effect of the loop contributions is a small shift of the allowed region in the parameter space towards lower values of $s_A$.

Finally, we have also included the effect of the heavy partners of all the fermions of the SM in
the loop. For that, we have considered several embeddings for the composite fermions. We have studied MCHM$_5$, where one introduces four composite quarks for each generation in the following representations of SO(5): $q^1,u^1\sim{\bf5}_{2/3}$ and $q'^1,d^1\sim{\bf5}_{-1/3}$. For the leptons we have used: $L^1,e^1\sim{\bf5}_{-1}$. We have also considered MCHM$_{10}$, with $q^1,u^1,d^1\sim{\bf10}_{2/3}$ and $L^1,e^1\sim{\bf10}_{-1}$~\cite{Contino:2006qr}. We also studied a set of representations that allows to solve the deviation in $A_{FB}^b$~\cite{Ciuchini:2014dea}, by embedding $q'^1\sim{\bf16}_{-5/6}$ and $d^1\sim{\bf4}_{-5/6}$~\cite{Andres:2015oqa}. For the sake of brevity, we discuss the corrections to the couplings of Eqs.~(\ref{eqn:effCoup}) and (\ref{eqn:effCoup2}) in the Appendix.

We find that the results shown in Fig.~\ref{contour-ratio-loop} remain almost unchanged when we include these new heavy resonances, and thus the effect on the allowed regions of the parameter space are negligible, further indicating the stability of the tree-level predictions. Moreover, since the effect is very small for the selected region of the parameter space, although the different representations contain different numbers of fermions and mild variations in the mixing, the results are practically independent of this choice.



\section{Discussion and predictions}\label{sec:predictions}

\begin{figure}
\begin{center}
\includegraphics[width=0.46\textwidth]{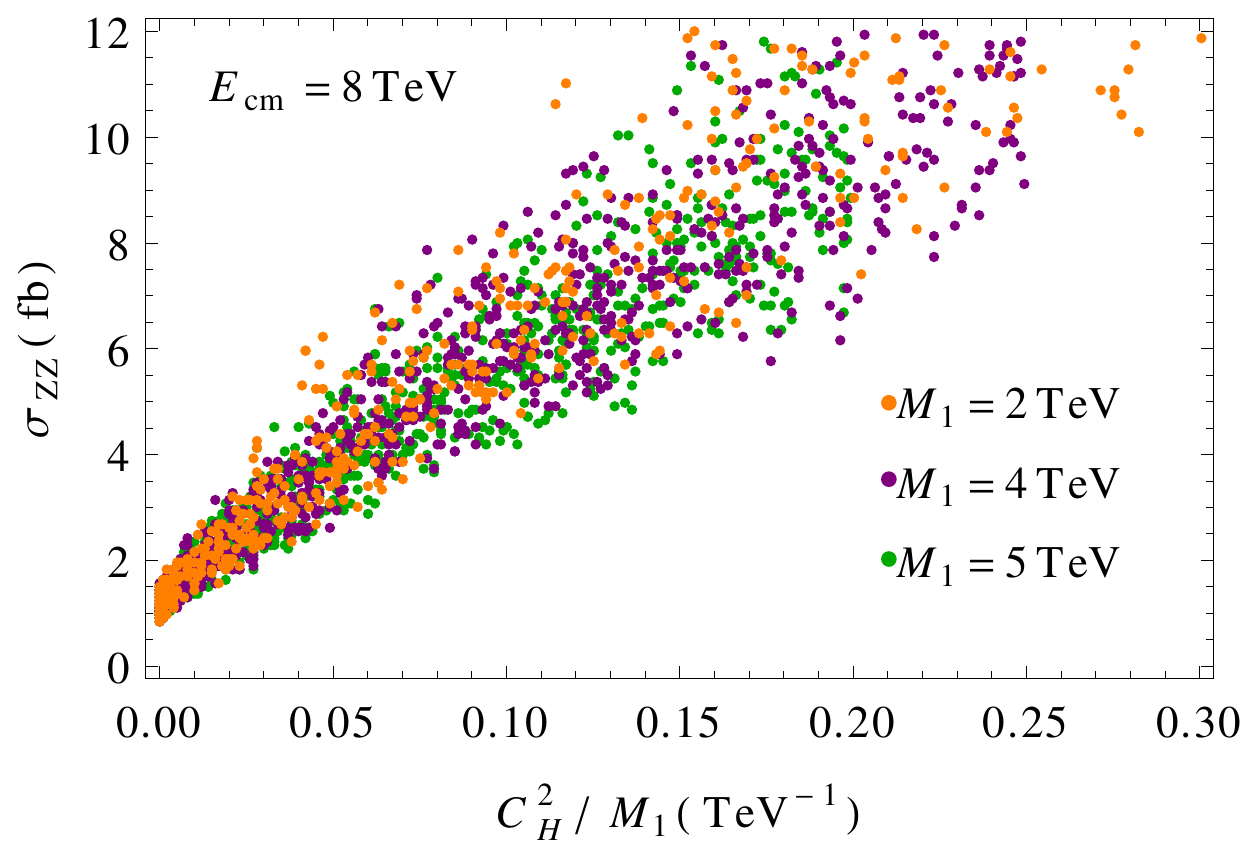}~
\includegraphics[width=0.45\textwidth]{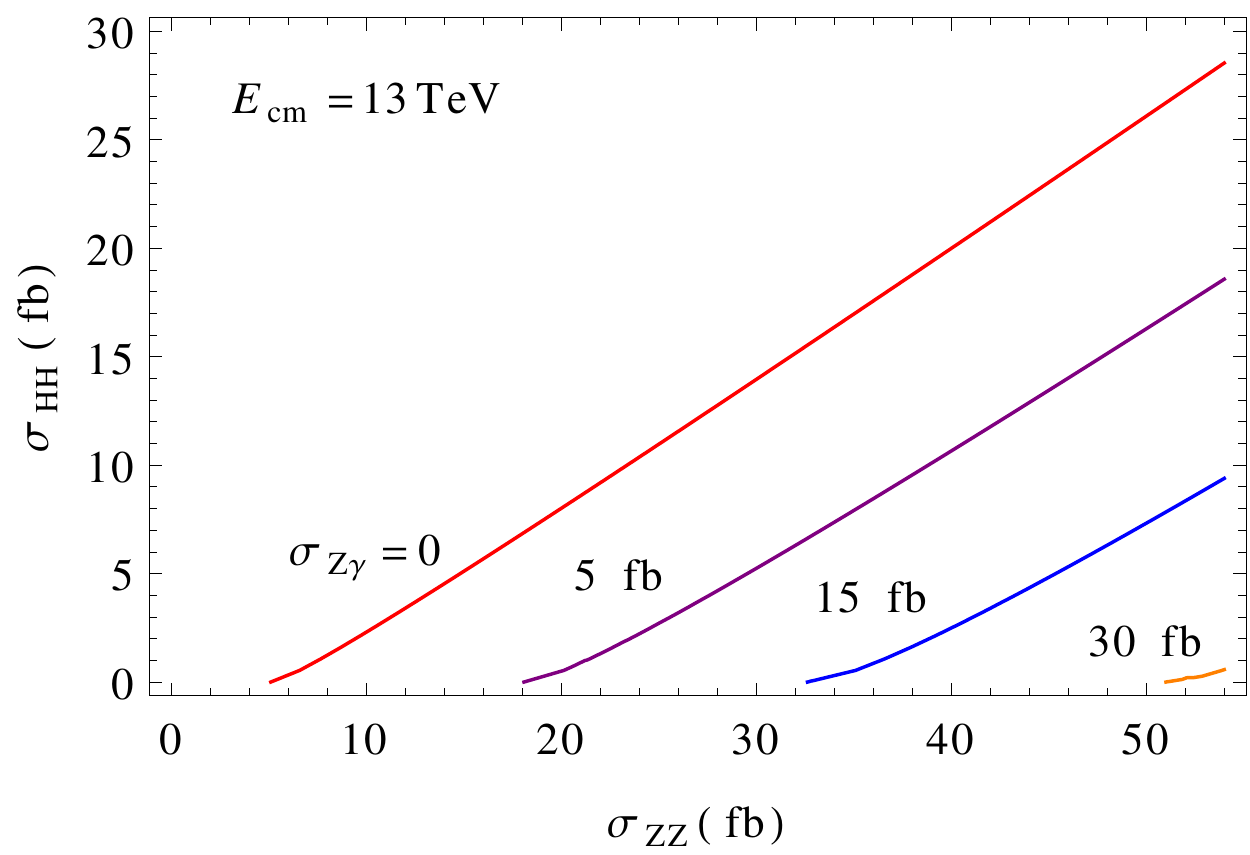}
\caption{
\small
Left: The $8\text{ TeV}$ production cross section for $ZZ$, as a function of $C_H^2/M_1$, for all the points in the scan that fulfill all the constraints and reproduce the $\gamma\gamma$ excess within $1\sigma$, for different values of $M_1$.  A similar distribution is obtained for the 13 TeV case, but with larger cross-sections.  Right: The model predicts $\sigma_{HH}$ as a function of $\sigma_{ZZ}$ and $\sigma_{Z\gamma}$, in this plot we show some contour plots for possible values of $\sigma_{Z\gamma}$.  Observe that this prediction is independent of $M_1$.
}
\label{plot:xsZZ}
\end{center}
\end{figure}

One of the main results in this work is that as $C_H$ increases then the gravity scale of site-1, $M_1$, should also increase in order to explain the data with the model (see Fig.~\ref{scatter} and discussion in text).  Since we found that the limit in increasing $C_H$ is the bound in $\sigma_{ZZ}=\sigma \times $BR$(ZZ)$, and that this limit is higher as $M_1$ increases, then we inquire which is the dependence of this quantity on $C_H$ and $M_1$.  We found that $\sigma_{ZZ}$ is approximately linear in $C_H^2/M_1$, as we plot in Fig.~\ref{plot:xsZZ}a.  

As previously discussed, in Fig.~\ref{scatter} the main limit to the allowed parameter space --besides $\sigma_{\gamma\gamma}$-- is $\sigma_{ZZ}$, and close by also $\sigma_{WW}$.  Other limits, as for instance $\sigma_{HH}$ and $\sigma_{tt}$ are not so tight.  For instance, green points in Fig.~\ref{scatter} for $M_1=2$ TeV have $\sigma_{HH}$ and $\sigma_{tt}$ less than $7$ fb and $170$ fb, respectively, for $\sqrt{s}=8$ TeV, whereas their respective limits are $39$ fb and $450$ fb.

Although we have included some loop corrections and verified that the results are only slightly modified, it would be desirable to include other corrections as well.  Observe that corrections to $C_{g,\gamma}^{\text{eff}}$ due to gauge boson loops can be self-consistently neglected since in our solutions we obtain that tree couplings between graviton and gauge bosons, $s_A^2/M_1^2$, are smaller than couplings between graviton and relevant SM fermions, $s^2_{q, t, b}/M_1^2$, and KK fermions. 
Modifications to $C_{Z,W}^{\text{eff}}$ are harder to compute, because there are different fermions running in the loop, and $C_H^{\text{eff}}$ also has to be computed at the same level.
The calculation of these effects is left for future work.

The model in this work also predicts graviton production through $WW$/$ZZ$ fusion, leading to a final state with two additional forward jets.  The strength of this production mechanism depends on $s_A^2$ and on $C_H$.  The larger is $C_H$, the stronger the longitudinal modes of $W$ and $Z$ couple to the graviton and the stronger is the production through this mechanism.  We have verified numerically using MadGraph \cite{Alwall:2014hca} that graviton production through vector boson fusion accounts for less than a percent than through gluon fusion.

Along this work we have assumed universal couplings to all gauge bosons, $s_A^2/M_1^2$.  This is, however, a simplification that is not actually required by the model to fit the data.  If this assumption is not satisfied, we may then expect to have some signal in $Z\gamma$, as well as changing the current limits in Fig.~\ref{scatter}.  Moreover, the cross-section in the $Z\gamma$ channel could be used to parameterize the departure of universality of the gauge boson couplings in order to formulate predictions within the model.

One of the most important predictions of the model is the relationship between the different decay channels.  In particular, we have checked that in most cases the next channel that would be seen is $ZZ$ and later $WW$, this is supposing that the sensitivities in the 13 TeV searches will have fairly the same ratios as in the 8 TeV searches.  Another prediction is that if the $Z\gamma$ channel is not seen --that is, that gauge couplings are universal-- then $\sigma_{HH}$ can be easily computed as a function of $\sigma_{\gamma\gamma}$ and $\sigma_{ZZ}$ using Eqs.~(\ref{zz}), (\ref{aa}) and (\ref{hh}).  Moreover, even if $\sigma_{Z\gamma} \neq 0$, one can predict the relationship for these three channels by using also Eq.~(\ref{za}), see Fig.~\ref{plot:xsZZ}b; where we have used the central value for $\sigma_{\gamma\gamma}$ and assumed $|C_Z|>|C_\gamma|$ which is expected if $\sigma_{ZZ}$ is measured in the near future.  All these predictions dispense of the value of $M_1$. 

An interesting feature of this model is that through Eq.~(\ref{eq:XtoKKandSM}) it predicts the decay of a composite particle into a graviton and a SM particle.
It would be interesting to explore phenomenologically this possibility.

As already mentioned in section \ref{sec:pheno}, the model generically predicts a relatively narrow width for the diphoton resonance.
If the width is found to be greater than $\sim 1\%$ of the mass, then the model would be in tension and should be adapted, probably with new decays.
It is worth pointing out that even in this case the relation between $\sigma_{ZZ}$ and $\sigma_{HH}$ presented in Fig.~\ref{plot:xsZZ}b holds.

There are also other general predictions which are only due to the spin of the resonance or the pNGB nature of the Higgs.  For instance, being the resonance a graviton created mainly through gluon fusion, it produces forward photons. There is also a very interesting phenomenology associated to the pNGB nature of the Higgs, that has been extensively studied. In most of the cases, a natural light Higgs requires the presence of new light fermions with masses of order TeV, usually called custodians, that could be produced and detected at LHC~\cite{Contino:2008hi,DeSimone:2012fs}. Another very important signal is the double Higgs production~\cite{Contino:2010mh,Contino:2012xk}.

\section{Conclusions}\label{sec:conclusions}
Along this work we have addressed the phenomenology of the $750$ GeV diphoton resonance by introducing a new spin-two massive state. A natural dynamics for this new state is to assume that there is a strongly interacting sector beyond the SM that generates resonances at the TeV scale. We have given a simplified description of the new sector in the framework of a 2-site model, where the first site contains the SM, the second site contains the first level of states of the strong dynamics, and a set of link fields allow interactions between the two sites. In this picture the diphoton resonance of spin-two corresponds to a massive graviton.
We have shown that, if the Higgs boson is a pseudo Nambu-Goldstone boson arising from the strongly interacting sector, one can avoid too large couplings between the massive graviton and the longitudinal polarization of the massive gauge bosons. This results in a natural way to avoid stringent bounds on $ZZ$ and $WW$ production from 8 TeV searches. In addition, the whole mechanism provides a solution to the hierarchy problem and a natural trigger for EWSB.
A possible UV-completion can be obtained by a five-dimensional theory in a warped Randall-Sundrum spacetime.


We have analyzed the phenomenology of the available observables within the framework of the proposed model and understood in detail which regions of parameter space are affected by each observable.  We have studied how the different variables in the model should be adapted to address each observable.  We have determined that the relevant variables in this model are the mixing angle for the vector bosons ($s_A$) that is determined by the ratio of couplings of the SM and the strongly interacting sector, the graviton coupling to the Higgs ($C_H$) and the gravity scale of the strongly interacting sector ($M_1$).  We have found the region in parameter space of these variables that correctly reproduces the observables and constraints.  

We have found that the production of diphotons and the constraints on the other channels lead to an interesting correlation between the Higgs coupling to the massive graviton and the the gravity scale in the composite sector. For $C_H=1$ the scale $M_1$ has to increase up to $4-5$~TeV, whereas  for smaller couplings: $C_H\sim0.5$, a smaller scale is allowed: $M_1\sim 1-2$~TeV. This is one of the main results of our work.

Since a new composite sector with an extended EW gauge symmetry may contain large multiplets of partners of the SM particles, we have estimated the size of the corrections at one-loop. For this purpose we have computed the contributions to the relevant graviton couplings by the presence of the SM third generation of quarks and the new fermionic sector. We have found that our results are stable upon these corrections, and also stable upon different embeddings of the fermions under the extended symmetry.

There are many interesting predictions, one of the most distinguishable is the relationship between $\sigma_{ZZ}$, $\sigma_{HH}$ and $\sigma_{Z\gamma}$ cross-sections at the 750 GeV resonant production scale, which is stated in Fig.~\ref{plot:xsZZ}b.  We expect that if $\sigma_{ZZ}$ is measured and $\sigma_{Z\gamma}$ is either measured or constrained, then the model would predict the value of $\sigma_{HH}$.


\section*{Acknowledgments}

We are grateful to Da Huang for useful communications concerning the NLO effects. We thank Gonzalo Torroba for useful discussions. This work was partially supported by ANPCyT PICT 2013-2266.

\section*{Appendix}\label{sec:appendix}
\renewcommand{\theequation}{{\rm{A}}.\arabic{equation}}
\setcounter{equation}{0}

We discuss here in more detail the calculation of the loop contributions presented in section \ref{sec:loop}.
Following the results of Ref.~\cite{Geng:2016xin}, the sum of the tree-level and loop induced amplitudes for $X^*\to \gamma\gamma$ can be accounted for with the following replacement of the bare coupling $C_\gamma^0$,
\begin{equation}
C_\gamma^0 \to C_\gamma^0 + \f{\alpha}{2\pi}\sum_i Q_i^2 C_i \f{N_c}{3} \left(
\f{2}{\epsilon}-\gamma_E+\ln(4\pi)+
A_G(\tau_i,\mu) 
\right)\,,
\end{equation}
where the sum runs over every fermion coupling to the graviton, and we have different terms for the right and left-handed ones.
Here $Q_i$ stands for the charge of the fermion in units of the positron charge $e$, $C_i$ represents its coupling to the graviton field and $N_c$ the number of colours. 
We work in dimensional regularization with $D = 4-\epsilon$.
Renormalizing the Wilson coefficient as
\begin{equation}
C_\gamma = C_\gamma^0
+ \f{\alpha}{2\pi}\sum_i Q_i^2 C_i \f{N_c}{3} \left(
\f{2}{\epsilon}-\gamma_E+\ln (4\pi) 
\right)\,,
\end{equation}
we are left with an effective coupling
\begin{equation}
C_\gamma^{\text{eff}}(\mu) =
C_\gamma + 
\f{\alpha}{2\pi}\sum_i Q_i^2 C_i \f{N_c}{3}
A_G(\tau_i,\mu)\,.
\end{equation}

For the SM resonances, for which $\tau_i = 4 m_i^2/m_X^2$ is always lower than 1, the loop function $A_G$ takes the form  \cite{Geng:2016xin}
\begin{eqnarray}
A_G(\tau,\mu) &=&  -\f{1}{12}\bigg[
-\f{9}{4}\tau(\tau+2) [2 \tanh^{-1}(\sqrt{1-\tau}) - i\pi]^2
\\
&+& 3(5\tau+4)\sqrt{1-\tau}[2 \tanh^{-1}(\sqrt{1-\tau}) - i\pi]
- 39\tau - 35 -12 \ln\f{\mu^2}{m_i^2} \bigg]\,.\nonumber
\end{eqnarray}
For the phenomenological results, we choose the renormalization scale to be $\mu = m_X$, given that it represents the typical energy scale of the process.
In particular, this choice yields a finite result for $A_G$ in the $\tau\to 0$ limit.

The situation for the heavy partners of the SM fermions is different, since we always have $\tau > 1$.
The extension of the previous result is simply
\begin{eqnarray}
A_G(\tau,\mu) &=& -\f{1}{12}\bigg[
\f{9}{4}\tau(\tau+2) [2 \tan^{-1}(\sqrt{\tau-1}) - \pi]^2
\\
&-& 3(5\tau+4)\sqrt{\tau-1}[2 \tan^{-1}(\sqrt{\tau-1}) - \pi]
- 39\tau  - 35 -12 \ln\f{\mu^2}{m_i^2} \bigg]\,.\nonumber
\end{eqnarray}
The suitable scale choice is, however, different from the previous case. The scale is imposed now by a matching condition between  the effective theory we are considering, which is valid up to a cut off $\Lambda \sim M_1$, and the full theory above this scale. Within this framework, the renormalization scale is set at the mass of the heavy fermions in order to avoid large logarithms in the Wilson coefficients as it is the usual procedure. Besides, we are not including running effects from $\mu = m_{\psi_1} \sim {\cal O}$(TeV) to $\mu = m_X \simeq 750$ GeV since we expect them to give minor corrections to already small loop contributions. 

The loop-induced modifications to the coupling of the graviton to a pair of gluons have the same features described above for the photon case.
The resulting expression for the effective coupling reads
\begin{equation}
C_g^{\text{eff}}(\mu) = C_g 
+ 
\f{\alpha_S}{2\pi}\f{1}{6}\sum_i C_i 
A_G(\tau_i,\mu)\,,
\end{equation}
where in this case the sum runs over each chirality of the fermions carrying colour.


\bibliography{biblio}

\end{document}